\documentclass[12pt]{article}
\usepackage{authblk}
\usepackage[bookmarksnumbered, colorlinks, plainpages]{hyperref}
\usepackage{amsmath, amsthm, amscd, amsfonts, amssymb, graphicx, color, booktabs}

\usepackage{cite}
\usepackage{subfigure}

\numberwithin{equation}{section}
\definecolor{email}{rgb}{0.00,0.00,0.84}
\begin{document}
\setcounter{page}{1}

\title{\large \bf 12th Workshop on the CKM Unitarity Triangle\\ Santiago de Compostela, 18-22 September 2023 \\ \vspace{0.3cm}
\LARGE Rare Charm Decays at BESIII}

\author[1]{Zhijun Li\footnote{lizhj37@mail2.sysu.edu.cn}}
\author[1]{Zhengyun You\footnote{youzhy5@mail.sysu.edu.cn}}
\affil[1]{School of Physics, Sun Yat-sen University, Guangzhou 510275, China}
\maketitle

\begin{abstract}
The rare and forbidden processes within the Standard Model offer an opportunity to explore potential new physics beyond the SM.  We summarize the research method and the recent results of rare charm decays at BESIII based on the extensive data samples in the  $\tau-c$ energy region, many of which impose stringent constraints on the new physics.
\end{abstract} \maketitle

\section{Introduction}

\noindent The Standard Model (SM) of particle physics has achieved great success, but it still has some puzzles, such as dark matter, matter and anti-matter asymmetry, fermion mass hierarchy, and more. There must be something new beyond the SM, which is referred to as new physics. The rare and forbidden processes within the SM could provide a golden opportunity to search for new physics, as they are not disturbed by the SM, and any signals that exceed expectations could indicate significant new physics phenomena. 

Beijing Spectrometer III (BESIII)~\cite{Ablikim:2009aa,Huang:2022wuo} is a general-purpose spectrometer for $\tau$-charm physics study in the center-of-mass energy range from 2.0 to 4.7~GeV. BESIII records symmetric $e^+e^-$ collisions provided by the Beijing Electron Positron Collider II (BEPCII) storage ring~\cite{Yu:IPAC2016-TUYA01} and has collected large data samples in this energy region~\cite{BESIII:2020nme}, such as 10 billion $J/\psi$ events, 2.7 billion $\psi(2S)$ events, $8~\rm{fb}^{-1}$ data at 3.773 GeV and more than $20~\rm{fb}^{-1}$ data above 4.0 GeV in total. With these extensive data samples at BESIII, it is possible to probe the new physics in the rare charm decays.

\section{BESIII rare charm decay measurements}

\subsection{Charmonium weak decays}
\begin{figure*}[htbp] \centering
	\setlength{\abovecaptionskip}{-1pt}
	\setlength{\belowcaptionskip}{10pt}

        \subfigure[]
        {\includegraphics[width=0.32\textwidth]{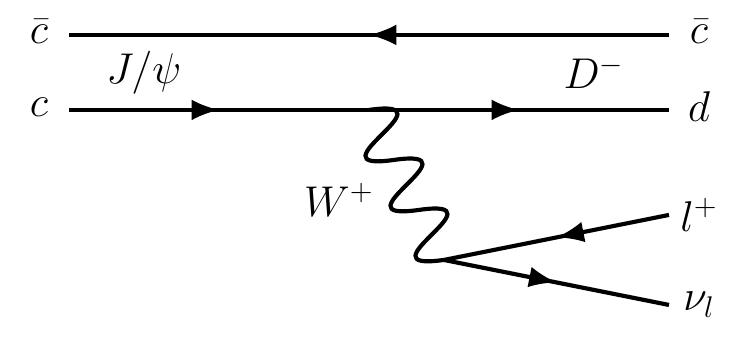}}
        \subfigure[]
        {\includegraphics[width=0.32\textwidth]{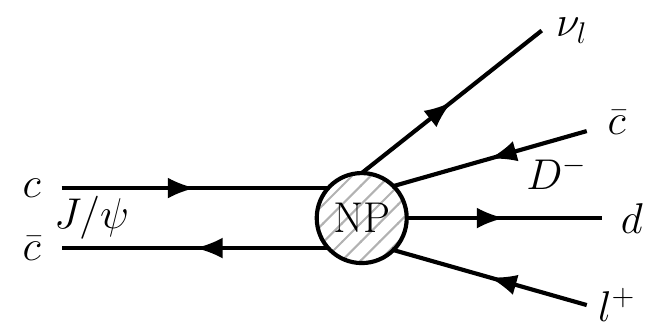}}
        \subfigure[]
        {\includegraphics[width=0.32\textwidth]{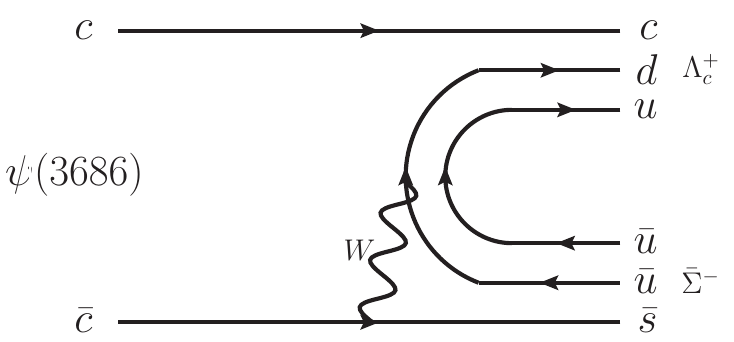}}\\

        \subfigure[]
         {\includegraphics[width=1.0\textwidth]{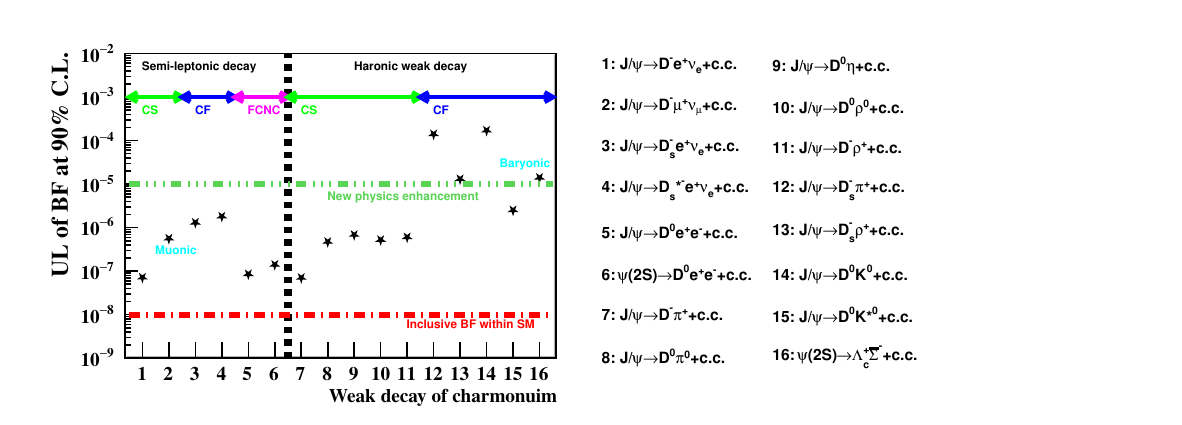}}\\
        
	\caption{
        The Feynman diagram of $D^{-}l^{+}\nu_{l}$ within the SM (a) and beyond the SM (b). (c) The Feynman diagram of $\psi(2S)\to\Lambda^+_c\bar{\Sigma}^-$ within the SM.
        (d) The current experimental constraints of the charmonium weak decays, where CS is Cabibbo-Suppressed, CF is Cabibbo-Favoured and FCNC is flavor changing neutral current.
        } 
	\label{fig:weak}
\end{figure*}

Charmonium weak decays are allowed in the SM, as shown in Figure~\ref{fig:weak} (a)(c), but highly suppressed by strong and electromagnetic interaction. The inclusive $J/\psi$ weak decay branching fraction(BF) is predicted to be in the order of $10^{-8}$ or below within the SM, and the exclusive BF is mostly in the order of $10^{-9}$ to $10^{-11}$. Some new physics models can enhance the BF of $J/\psi$ weak decay to $10^{-5}$ like Figure~\ref{fig:weak} (b). Based on $(10087\pm44)\times 10^{6}$ $J/\psi$ events collected at BESIII, the semi leptonic decay $J/\psi\to D^{-}l^{+}\nu_{l}+c.c.$ ($l=e,\mu$) is studied. 
No significant signal is found,
and the upper limit(UL) of the BF is set to $\mathcal{B}(J/\psi\to D^{-}e^{+}\nu_{e}+c.c.)<7.1\times10^{-8}$ and $\mathcal{B}(J/\psi\to D^{-}\mu^{+}\nu_{\mu}+c.c.)<5.6\times10^{-7}$ at 90\% confidence level(CL). For the electron channel, its UL is improved by more than two orders of magnitude~\cite{BESIII:2021mnd}, and for the muon channel, it is the first search for the weak interaction between muon and $J/\psi$~\cite{BESIII:2023fqz}.

\vspace{-0.0cm}
\vspace{-0.0cm}

For the charmonium bound state $J/\psi$, it can not decay to the lightest charm baryon $\Lambda^+_c$ within the SM due to the baryon number conservation and the mass limit. But for the excited state $\psi(2S)$, it is possibly decayed to $\Lambda^+_c\bar{\Sigma}^-+c.c.$ (Figure~\ref{fig:weak} (c)) and it is also studied at BESIII based on $(448.1\pm2.9)\times10^6$ $\psi(2S)$ events. 
The UL of its BF is $1.4\times 10^{-5}$ at 90\% CL, which is the first search for the charmonium baryonic weak decay~\cite{BESIII:2022ibp}. The current experimental constraints of the charmonium weak decays are summarized in Figure~\ref{fig:weak} (d).

\subsection{Flavor changing neutral current processes}
\begin{figure*}[htbp] \centering
	\setlength{\abovecaptionskip}{-1pt}
	\setlength{\belowcaptionskip}{10pt}
 
        \subfigure[]
        {\includegraphics[width=0.24\textwidth]{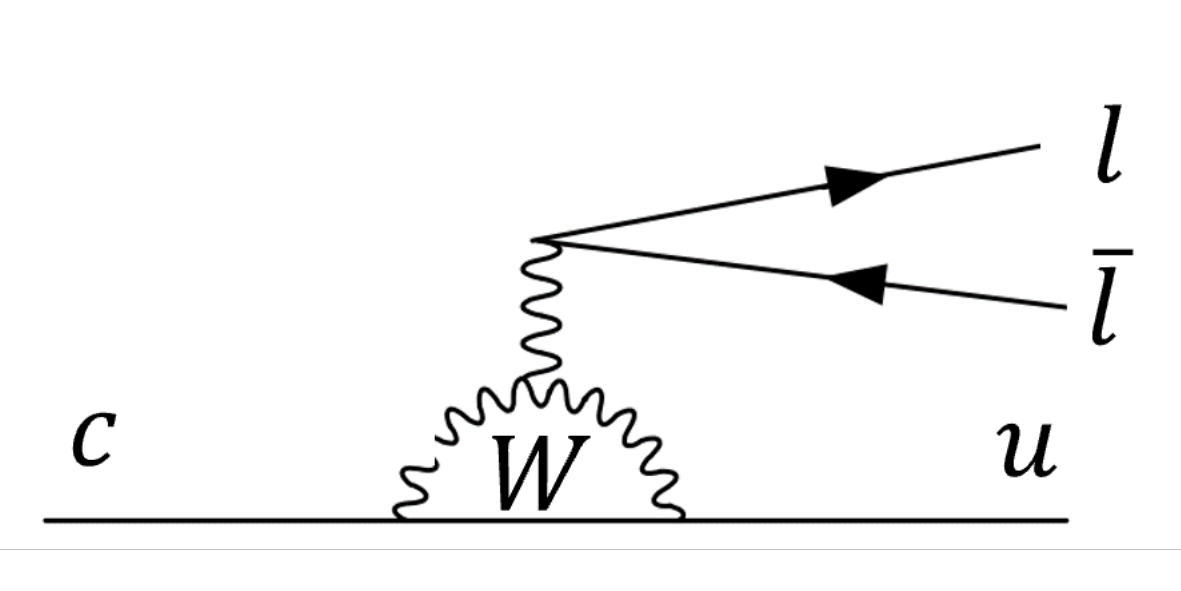}}
        \subfigure[]
        {\includegraphics[width=0.24\textwidth]{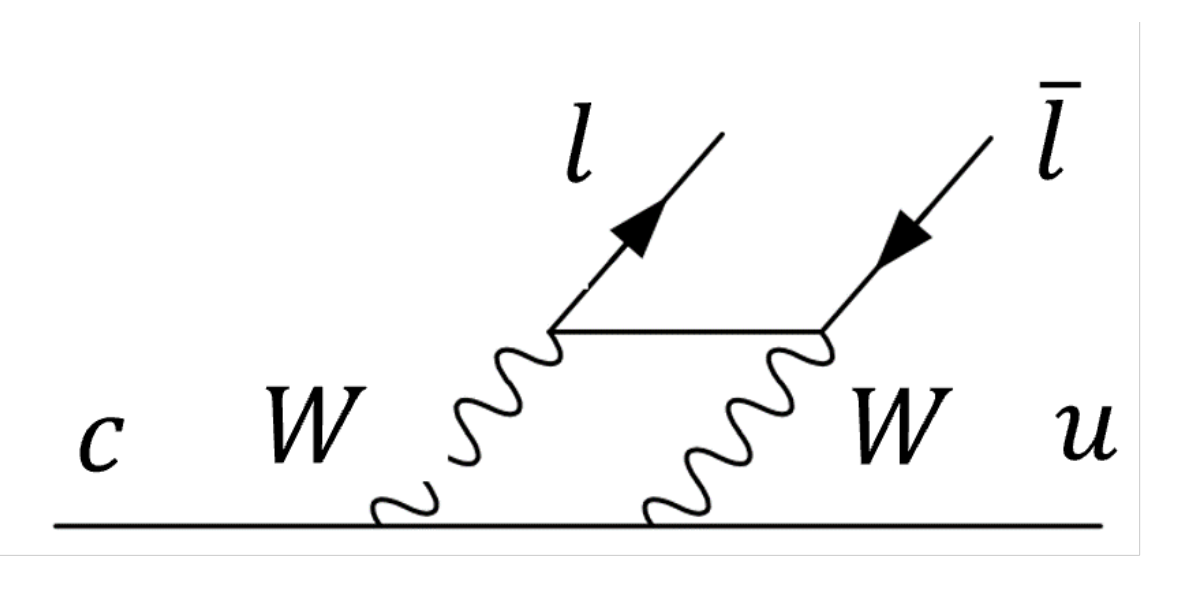}}
        \subfigure[]
        {\includegraphics[width=0.24\textwidth]{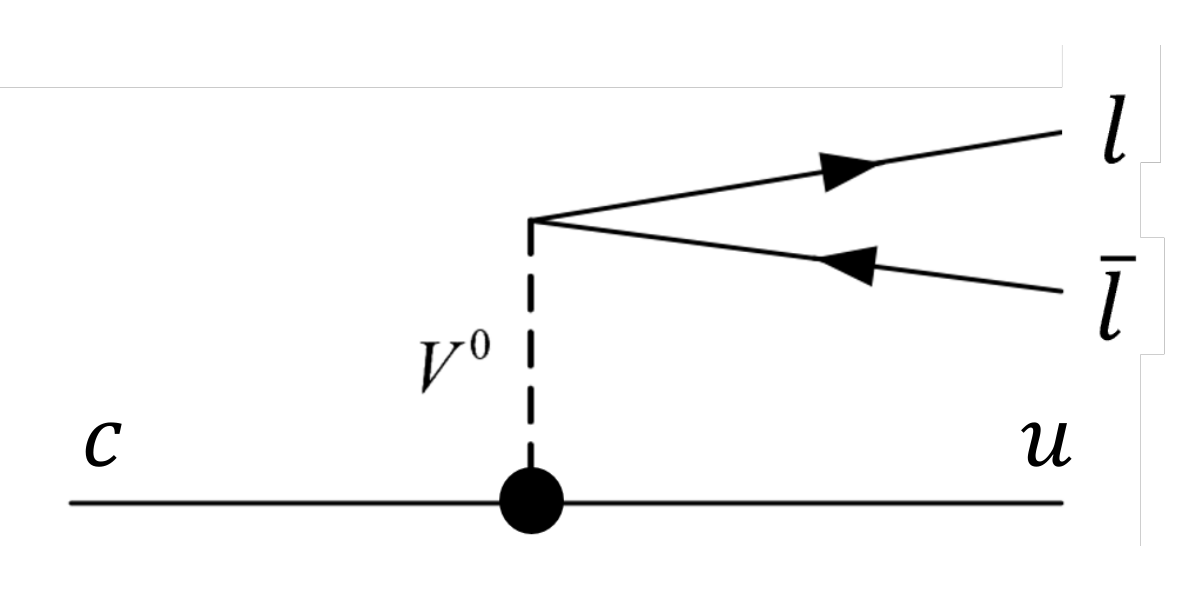}}
         \subfigure[]
         {\includegraphics[width=0.24\textwidth]{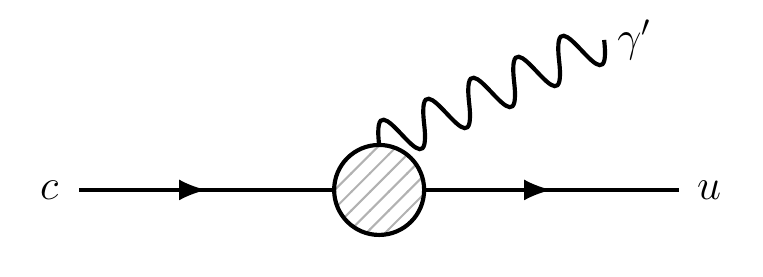}}\\
        
	\caption{(a)(b) The Feynman diagram of FCNC process of $c\to u l\bar{l}$ ($l=e,\mu,\nu$). (c) The LD contribution of $c\to u l\bar{l}$ ($l=e,\mu$). (d) The $cu\gamma'$ coupling in the dimension-six operator beyond the SM. 
        } 
	\label{fig:fcnc}
\end{figure*}

Due to the Glashow-Iliopoulos-Maiani (GIM) mechanism, flavor changing neutral current (FCNC) processes are strongly suppressed in the SM, which are forbidden at the tree level and can only happen through a loop diagram. The SM predictions for the BFs of the FCNC processes of the charm quark would not exceed the level of $10^{-9}$. The main channel to study the charm FCNC process is $c\to u l\bar{l}$ ($l=e,\mu,\nu$), as shown in Figure~\ref{fig:fcnc} (a) (b), although it usually also has the non-FCNC contribution from the long-distance (LD) by vector meson such as $\rho$ meson (Figure~\ref{fig:fcnc} (c)). Based on $2.93~\rm{fb}^{-1}$ data samples at $\sqrt{s}=3.773$ GeV, a series of the FCNC processes $D\to h(h^{(')})e^+e^-$ is studied, where $h$ means hadron, such as $D^0\to\pi^0e^+e^-$ and $D^+\to K^0_S\pi^+e^+e^-$. The double tag method is applied with clean background, and the UL of these BFs are set to $10^{-5}$ to $10^{-6}$ at 90\% CL~\cite{BESIII:2018hqu}. Except for the charged lepton channels, the neutrino channel $D^0\to\pi^0\nu\bar{\nu}$ with no LD contribution has also been studied for the first time recently. 
The BF of $D^0\to\pi^0\nu\bar{\nu}$ is less than $2.1\times10^{-4}$ at 90\% CL~\cite{BESIII:2021slf}.
\vspace{-0.0cm}
\vspace{-0.0cm}

Since the charm FCNC process is rare within the SM, it serves as a sensitive probe for searching for new physics that could also induce FCNC processes, such as the existence of a massless dark photon. The presence of an additional Abelian gauge group would give rise to a corresponding gauge boson, the dark photon. If the symmetry of this group remains unbroken, it would result in the existence of a massless dark photon $\gamma'$. $\gamma'$ has no direct interaction with the SM particles but can be coupled with the SM particles in a higher dimension operator from the new physics energy scale, such as the $cu\gamma'$ coupling shown in Figure~\ref{fig:fcnc} (d). The FCNC processes $\Lambda_c\to p \gamma'$ is searched with $4.5~\rm{fb}^{-1}$ data collected between $\sqrt{s}=4.6\sim4.699~\rm{GeV}$. 
Since there are no exceeding signals, the BF UL of $\gamma'$ production is set to $8.0\times 10^{-5}$ at 90\% CL~\cite{BESIII:2022vrr}.

\subsection{Lepton number and baryon number violation}
\begin{figure*}[htbp] \centering
	\setlength{\abovecaptionskip}{-1pt}
	\setlength{\belowcaptionskip}{10pt}
 
        \subfigure[]
        {\includegraphics[width=0.32\textwidth]{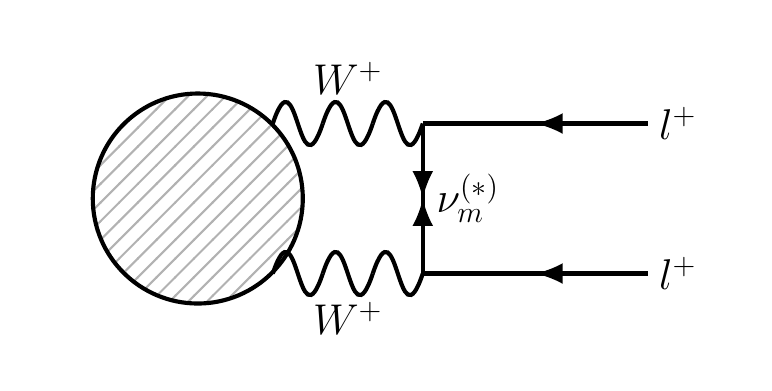}}
        \subfigure[]
        {\includegraphics[width=0.32\textwidth]{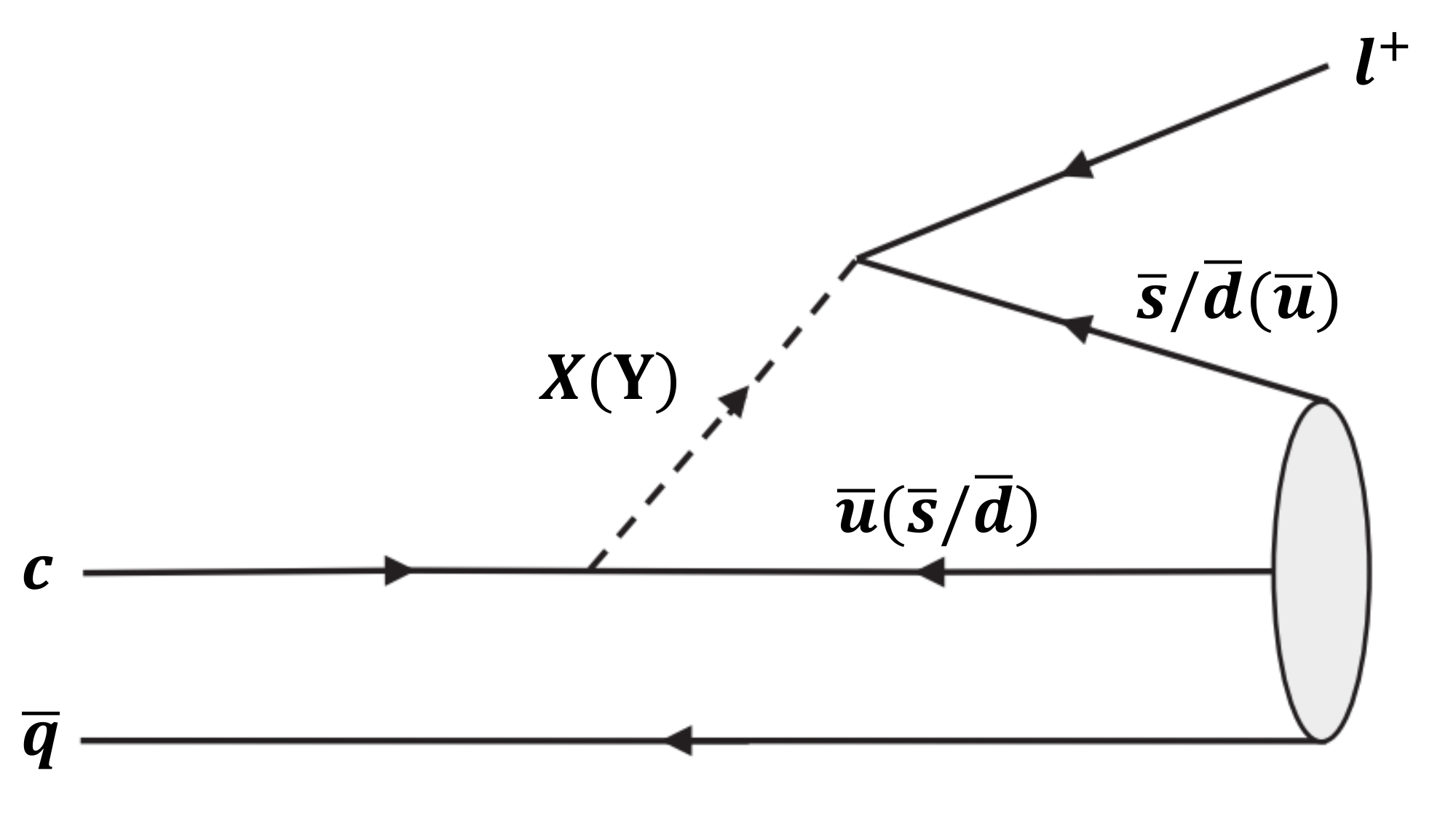}}
        \subfigure[]
        {\includegraphics[width=0.32\textwidth]{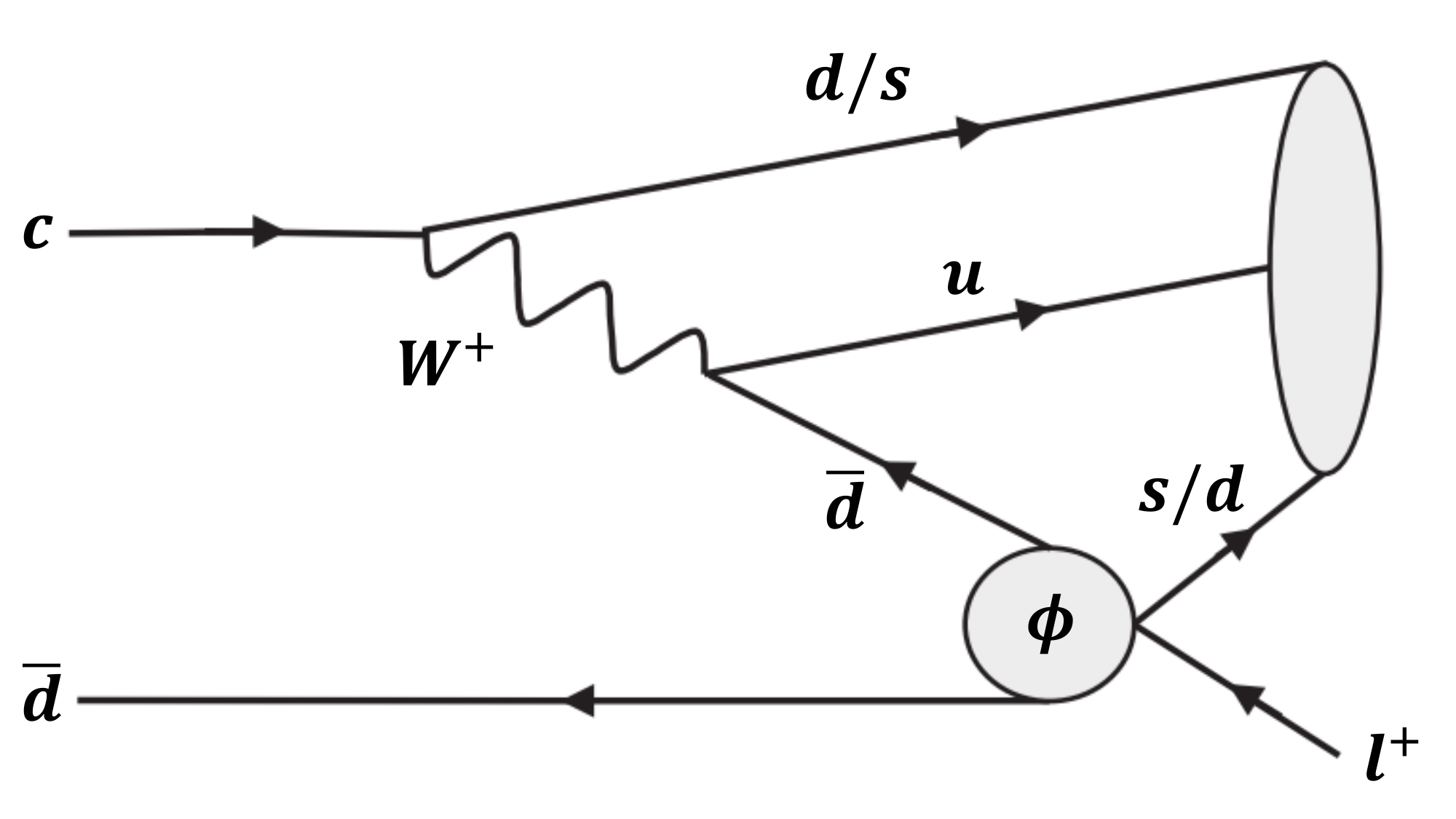}}\\
        \subfigure[]
        {\includegraphics[width=1.0\textwidth]{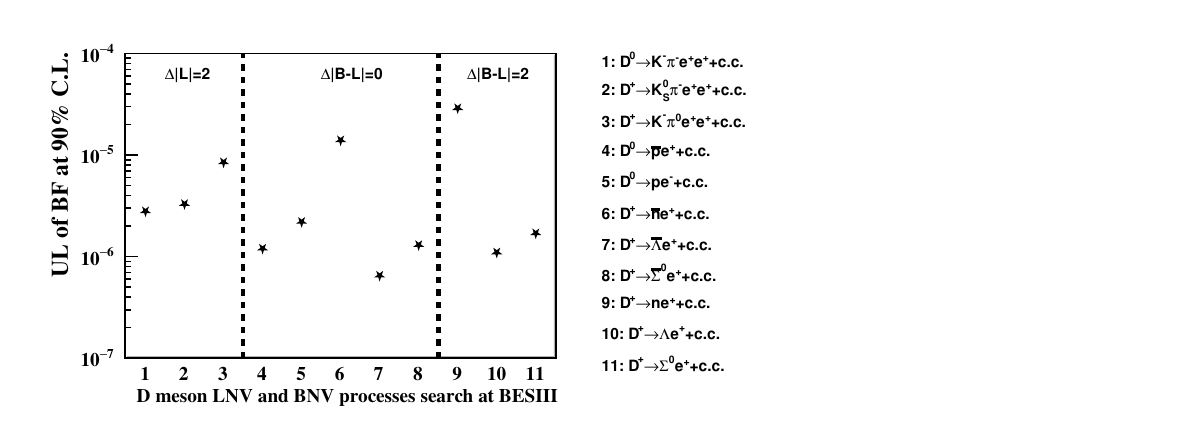}}
        
	\caption{(a) LNV process mediated by the Majorana neutrino. (b) BNV process of $D$ meson mediated by the leptoquarks. (c) BNV process of $D$ meson mediated by an elementary scalar field. (d) The summary of LNV and BNV process measurement of $D$ meson at BESIII.} 
	\label{fig:LBNV}
\end{figure*}

Lepton number (L) and baryon number (B) are always conserved within the SM, but there are some motivations to consider scenarios involving lepton number violation (LNV) and baryon number violation (BNV). The nature of neutrinos, specifically whether they are Dirac or Majorana particles, remains an open question. If neutrinos are Majorana particles, it could lead to the production of LNV processes with $\Delta |L| = 2$, as illustrated in Figure~\ref{fig:LBNV} (a). In the universe, the baryon anti-baryon number is highly asymmetric, it also indicates the potential existence of BNV. BNV processes can be generated through dimension-six operators with $\Delta|B-L| = 0$, mediated by leptoquarks X (with charge $\frac{4}{3}e$) or Y (with charge $\frac{1}{3}e$), or through dimension-seven operators with $\Delta|B-L| = 2$ mediated by an elementary scalar field $\phi$, as shown in Figure~\ref{fig:LBNV} (b) (c). Neutrinoless double beta decay (and proton decay) is the most promising avenue for exploring LNV (BNV). It is also meaningful to investigate these violations in heavy quark systems such as the $D$ meson.
Using the data sample of  $2.93~\rm{fb}^{-1}$ at $\sqrt{s}=3.773$ GeV, a search for a series of LNV and BNV processes of $D$ meson is conducted, setting the UL of their BFs in the range of $10^{-5}$ to $10^{-6}$~\cite{BESIII:2019oef,BESIII:2021krj,BESIII:2019udi,BESIII:2022svy}, as shown in Figure~\ref{fig:LBNV} (d).

\vspace{-0.0cm}
\vspace{-0.0cm}

\subsection{Charged lepton flavor violation}
\begin{figure*}[htbp] \centering
	\setlength{\abovecaptionskip}{-1pt}
	\setlength{\belowcaptionskip}{10pt}
 
        \subfigure[]
        {\includegraphics[width=0.32\textwidth]{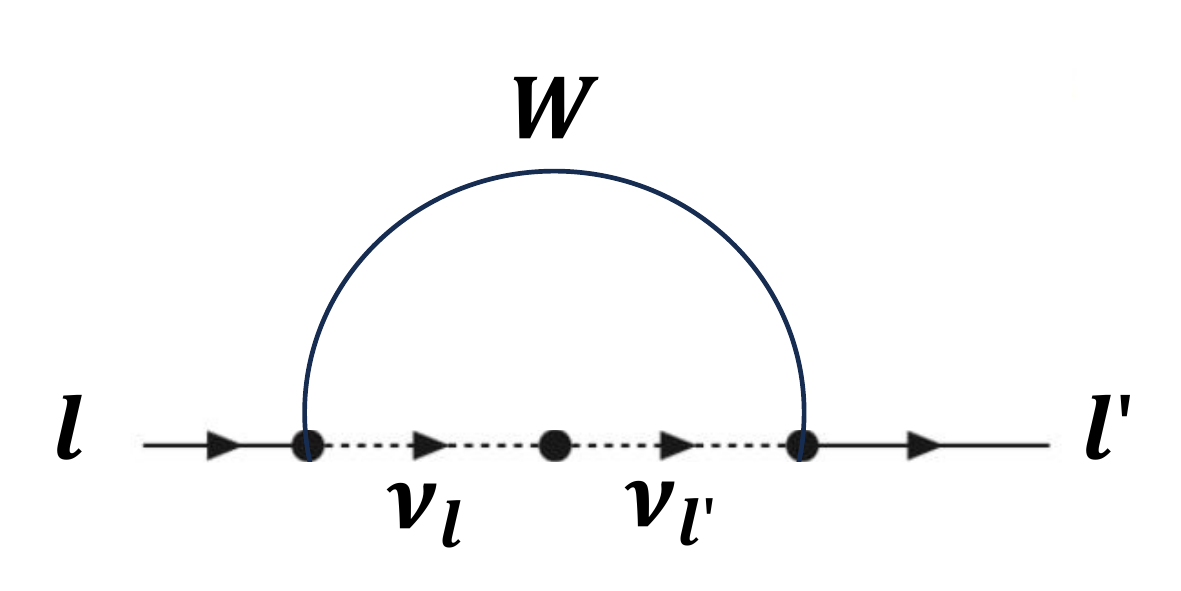}}
        \subfigure[]
        {\includegraphics[width=0.32\textwidth]{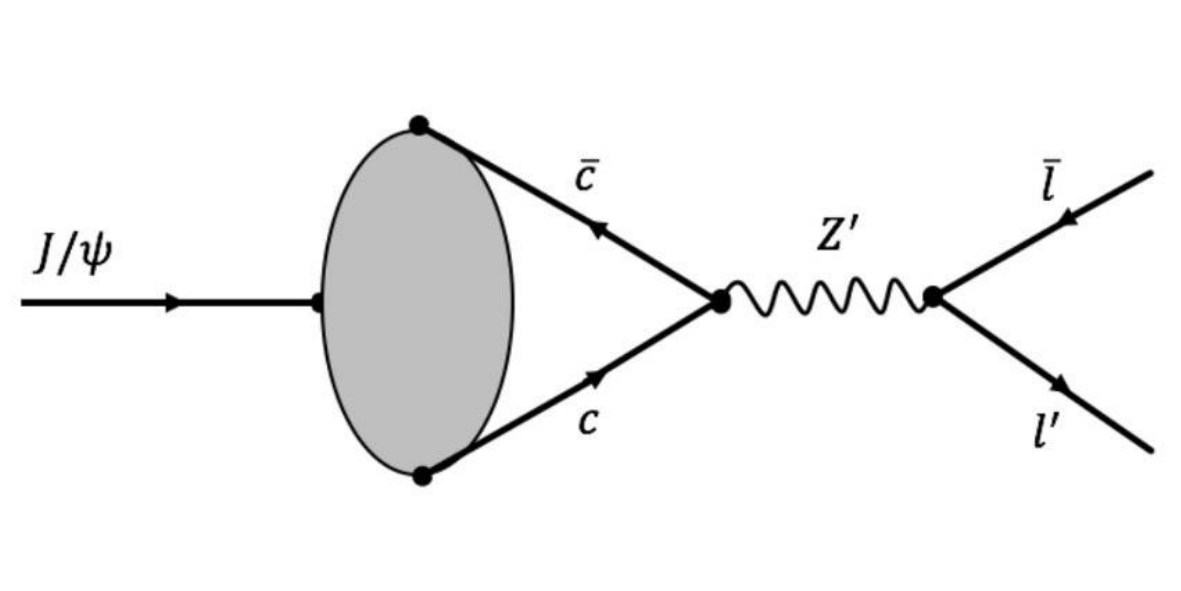}}
        \subfigure[]
        {\includegraphics[width=0.32\textwidth]{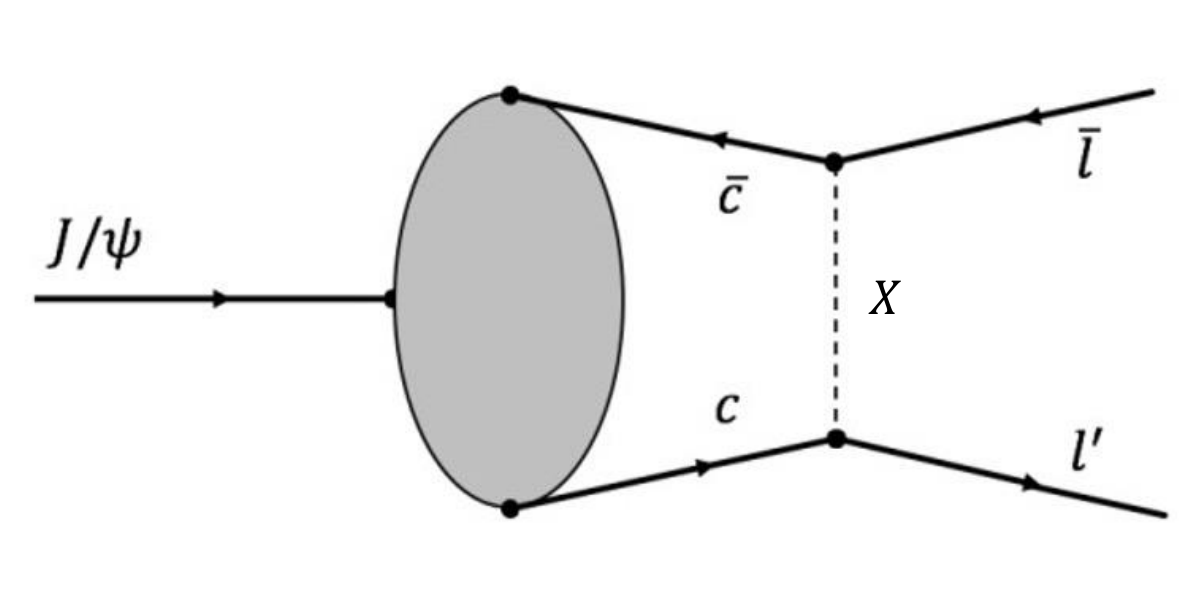}}\\
        \subfigure[]
        
	\caption{(a) CLFV within the extended SM. (b) CLFV mediated by $Z'$ particle. (c) CLFV mediated by leptoquark. 
        } 
	\label{fig:CLFV}
\end{figure*}

With the neutrino mixing, charged lepton flavor violation (CLFV) is allowed in the extended SM (Figure~\ref{fig:CLFV} (a)), but very rare to be detected. Some new physics particles can also allow the CLFV, such as $Z'$ particle and leptoquark (Figure~\ref{fig:CLFV} (b)(c)), which may enhance the BF of CLFV to a detectable level. Therefore, the discovery of any CLFV process would be a clear signal of new physics beyond the SM. The large $J/\psi$ samples of BESIII provide an opportunity to search CLFV in the heavy quark system, such as $J/\psi\to e\tau$ and $J/\psi\to e \mu$.  
Neither of the two searches found significant signals, and the BFs are set to be $\mathcal{B}(J/\psi\to e\tau)<7.5\times 10^{-8}$ and $\mathcal{B}(J/\psi\to e\mu)<4.5\times 10^{-9}$ at 90\% CL, which is the most precise CLFV search in heavy quarkonium system~\cite{BESIII:2021slj,BESIII:2022exh}. 

\vspace{-0.0cm}
\vspace{-0.0cm}

\section{Summary}
The investigation of rare decays and symmetry violation processes are essential for probing new physics beyond the SM. BESIII has conducted extensive studies on these processes in the charm system and has provided stringent constraints on them. In the future, BESIII will collect more data samples in $\tau-c$ region, such as $20~\rm{fb}^{-1}$ data samples at $\sqrt{s}=3.773~\rm{GeV}$ and new data samples at $\sqrt{s}\sim5~\rm{GeV}$. More comprehensive and improved results are expected in the near future.


{\bf Acknowledgements.} 
This work is supported in part by National Key Research and Development Program of China under Contracts Nos. 2020YFA0406400, 2020YFA0406300, 2023YFA1606000; Joint Large-Scale Scientific Facility Funds of the National Natural Science Foundation of China (NSFC) and Chinese Academy of Sciences (CAS) under Contracts Nos. U1932101.



\bibliographystyle{unsrt}
\bibliography{mybib}

\begin{thebibliography}{10}

\bibitem{Ablikim:2009aa}
M.~Ablikim et~al.
\newblock {Design and Construction of the BESIII Detector}.
\newblock {\em Nucl. Instrum. Meth. A}, 614:345--399, 2010.

\bibitem{Huang:2022wuo}
Kai-Xuan Huang et~al.
\newblock {Method for detector description transformation to Unity and
  application in BESIII}.
\newblock {\em Nucl. Sci. Tech.}, 33(11):142, 2022.

\bibitem{Yu:IPAC2016-TUYA01}
Chenghui Yu et~al.
\newblock {BEPCII Performance and Beam Dynamics Studies on Luminosity}.
\newblock In {\em {7th International Particle Accelerator Conference}}, 2016.

\bibitem{BESIII:2020nme}
M.~Ablikim et~al.
\newblock {Future Physics Programme of BESIII}.
\newblock {\em Chin. Phys. C}, 44(4):040001, 2020.

\bibitem{BESIII:2021mnd}
Medina Ablikim et~al.
\newblock {Search for the rare semi-leptonic decay $J/\psi\to
  D^{-}e^{+}\nu_{e}+c.c.$}.
\newblock {\em JHEP}, 06:157, 2021.

\bibitem{BESIII:2023fqz}
Medina Ablikim et~al.
\newblock {Search for the semi-muonic charmonium decay $J/\psi\to
  D^{-}\mu^{+}\nu_{\mu}+c.c.$}.
\newblock {\em JHEP}, 01:126, 2024.

\bibitem{BESIII:2022ibp}
Medina Ablikim et~al.
\newblock {Search for the weak decay $\psi(3686) \to \Lambda_c^{+}
  \bar{\Sigma}^- +c.c$}.
\newblock {\em Chin. Phys. C}, 47(1):013002, 2023.

\bibitem{BESIII:2018hqu}
M.~Ablikim et~al.
\newblock {Search for the rare decays $D\to h(h')e^+e^-$}.
\newblock {\em Phys. Rev. D}, 97(7):072015, 2018.

\bibitem{BESIII:2021slf}
M.~Ablikim et~al.
\newblock {Search for the decay $D^{0} \to \pi^{0} \nu \bar{\nu}$}.
\newblock {\em Phys. Rev. D}, 105(7):L071102, 2022.

\bibitem{BESIII:2022vrr}
M.~Ablikim et~al.
\newblock {Search for a massless dark photon in $\Lambda_c \to p \gamma'$
  decay}.
\newblock {\em Phys. Rev. D}, 106(7):072008, 2022.

\bibitem{BESIII:2019oef}
Medina Ablikim et~al.
\newblock {Search for heavy Majorana neutrino in lepton number violating decays
  of $D\to K \pi e^+ e^+$}.
\newblock {\em Phys. Rev. D}, 99(11):112002, 2019.

\bibitem{BESIII:2021krj}
Medina Ablikim et~al.
\newblock {Search for baryon- and lepton-number violating decays $D^0
  \rightarrow \overline{p} e^+$ and $D^0 \rightarrow pe^-$}.
\newblock {\em Phys. Rev. D}, 105(3):032006, 2022.

\bibitem{BESIII:2019udi}
Medina Ablikim et~al.
\newblock {Search for baryon and lepton number violating decays
  $D^+\to\bar\Lambda(\bar\Sigma^0)e^+$ and $D^+\to\Lambda(\Sigma^0)e^+$}.
\newblock {\em Phys. Rev. D}, 101(3):031102, 2020.

\bibitem{BESIII:2022svy}
M.~Ablikim et~al.
\newblock {Search for baryon and lepton number violating decay
  $D^\ensuremath{\pm}\rightarrow{}n(\bar{n})e^{\pm}$}.
\newblock {\em Phys. Rev. D}, 106(11):112009, 2022.

\bibitem{BESIII:2021slj}
Medina Ablikim et~al.
\newblock {Search for the charged lepton flavor violating decay $J/\psi\to
  e\tau$}.
\newblock {\em Phys. Rev. D}, 103(11):112007, 2021.

\bibitem{BESIII:2022exh}
Medina Ablikim et~al.
\newblock {Search for the lepton flavor violating decay~$J/\psi\to e\mu$}.
\newblock {\em Sci. China Phys. Mech. Astron.}, 66(2):221011, 2023.

\end{thebibliography}

\end{document}